# Coexisting Holes and Electrons in High-$T_C$ Materials: Implications from Normal State Transport

Dale R. Harshman[*,1,2,3], John D. Dow[3,4], and Anthony T. Fiory[5]

[1]Physikon Research Corporation, Lynden, WA 98264, USA
[2]Department of Physics, University of Notre Dame, Notre Dame, IN 46556, USA
[3]Department of Physics, Arizona State University, Tempe, AZ 85287, USA
[4]Institute for Postdoctoral Studies, 6031 East Cholla Lane, Scottsdale, AZ 85253, USA
[5]Department of Physics, New Jersey Institute of Technology, Newark, NJ 07102, USA

Abstract

Normal state resistivity and Hall effect are shown to be successfully modeled by a two-band model of holes and electrons that is applied self-consistently to (i) DC transport data reported for eight bulk-crystal and six oriented-film specimens of $YBa_2Cu_3O_{7-\delta}$, and (ii) far-infrared Hall angle data reported for $YBa_2Cu_3O_{7-\delta}$ and $Bi_2Sr_2CaCu_2O_{8+\delta}$. The electron band exhibits extremely strong scattering; the extrapolated DC residual resistivity of the electronic component is shown to be consistent with the previously observed excess thermal conductivity and excess electrodynamic conductivity at low temperature. Two-band hole-electron analysis of Hall angle data suggest that the electrons possess the greater effective mass.

## 1. Introduction

Studies of normal-state DC resistivity and Hall effect for cuprates based on Y or on (Bi,Sr) find nearly linear-in-T temperature dependences in the resistivity $\rho$ and the Hall number density $n_H = 1/e|R_H|$, where $R_H$ is the Hall coefficient and e is the elementary charge (positive, by convention) [1-4]. Discovery of these results in $YBa_2Cu_3O_{7-\delta}$ prompted early speculation of electron-hole compensation in the a-b plane transport to explain the near-linear temperature dependence of $n_H$, ascribing the positive sign and temperature dependence of $R_H$ to dominance of transport by holes over electrons with differing mobilities [1-4]. Although offered up hypothetically at the time, it was deemed unlikely that compensation of electron and hole mobilities would lead to $\rho$ and $n_H$ proportional to T and Hall mobility $\mu_H$ proportional to $T^{-2}$. Strict linearity in T for $\rho$ and $n_H$ yields a $T^2$ dependence for the cotangent of the Hall angle $\cot\theta_H$, which is the functional form against which numerous authors had tested their data [5-20]. Independent

---

studies of the Bi-cuprates have shown that the temperature dependence (of cot $\theta_H$) is not precisely $T^2$ and that it varies with doping [21,22]. We discuss in section 2 that the exponent similarly deviates from 2 in the case of $YBa_2Cu_3O_{7-\delta}$. For electrical transport along the *c*-axis of $YBa_2Cu_3O_{7-\delta}$, the resistivity is a decreasing function of temperature above $T_C$, while the Hall coefficient is negative and nearly independent of temperature [3,23]. These properties show the presence of marginally-metallic transport by negatively-signed mobile charge carriers.

For DC transport measurements this work focuses on the high-$T_C$ superconductor $YBa_2Cu_3O_{7-\delta}$ (the $T_C \sim 90$ K compound for $\delta \approx 0.05$), where results in the normal state reported over the years by numerous research groups have created a large data base with high reliability because of data redundancy. Since the AC Hall effect determines scattering rates and effective masses, we augment our analysis with frequency and temperature-dependence measurements of the Hall angle in the far infrared in $YBa_2Cu_3O_{7-\delta}$ and $Bi_2Sr_2CaCu_2O_{8+\delta}$, to obtain information unavailable from DC measurements alone.

Various early theoretical models have been proposed to explain the Hall effect. Markiewicz had based an electron-hole interpretation of the DC Hall effect on the peculiar assumption that the hole component of conductivity scales with the measured conductivity, which is unverified [24]. Several other authors have proposed interpretations in terms of two-band models [4], including where the signs of the carriers need not be opposite [25-27], percolation models [28], and single band models with variations in mobility [29]. Theoretical explanations that were advanced include the doped Mott insulator [30], Luttinger liquid theory [31], phenomenology of two transport relaxation times [32], and k-space variation or anisotropy in scattering rate [33,34,35-37]. For example, in Ref. [35] temperature dependence in $R_H$ arises from temperature-dependent anisotropy of the scattering length $l_k = v_k\tau_k$ (velocity-scattering-time product) around the Fermi contour, as discussed by Ong [37]. A different approach is taken in Ref. [36], where energy levels E above the saddle point energy $E_S$ of a single $CuO_2$ band are treated as hole-like and levels $E < E_S$ as electron-like; temperature dependence of $R_H$ is then contained in a ad-hoc model for energy dependence in scattering anisotropy.

In recent theoretical treatments it was argued that the Hall effect in the normal state of high-$T_C$ superconductors involves no new physics [38,39] beyond interpreting the linear-T $\rho(T)$ in the framework of marginal Fermi liquid phenomenology [40]. High frequency Hall effect measurements support this perspective, since the Hall scattering rate was found to be consistent with the normal scattering rate of electrical transport [41]. By determining inelastic scattering from $\rho(T)$ and by deducing anisotropic impurity scattering from angle-resolved photoemission spectroscopy (ARPES) and $\rho(T\rightarrow 0)$, the aforementioned theory offers an explanation for weaker-than $T^2$ temperature dependences observed in cot $\theta_H$ [39]. However, as we show in section 4, the non-zero extrapolated $\rho(T\rightarrow 0)$ and other signatures of hole impurity scattering as required by Ref. [39] are not always observed, whereas a temperature dependence in $R_H$ is, negating the plausibility of this theory. Introducing two carrier bands, one of electrons and one of holes, as presented in section 3, successfully averts a necessity for impurity-hole scattering.

We begin with a brief discussion of the DC data surveyed for this article in section 2. Our two-band model for the normal-state resistivity and the Hall coefficient is developed in section 3. The analysis methods employed and our findings for the DC and AC Hall effect are discussed in section 4. Relationships of these findings to the superconducting state are discussed in section 5 and concluding remarks are provided in section 6. We shall see that the resistivity and Hall effect can be successfully modeled with two bands of carriers; holes and electrons.

## 2. Survey of experimental DC results for $YBa_2Cu_3O_{7-\delta}$

In this work, we carefully examine DC experimental data for $\rho$ and $R_H$ as reported in the literature for eight bulk crystal samples [3, 6-8, 10,12,19] and six *c*-axis oriented thin films [12-17] of $YBa_2Cu_3O_{7-\delta}$ (with $T_C$'s $\geq 90$ K) in order to elucidate actual temperature dependences $\rho(T)$ and $n_H(T)$. All of the specimens surveyed pertain to nominally stoichiometric "90 K phase" $YBa_2Cu_3O_{7-\delta}$. The data treated here were obtained by transcribing published figures into digital form (with 0.2% full-scale accuracy of transcription). In addition to sample-dependent variations in their magnitudes, we also find various deviations from strict linearity in the temperature dependence of the resistivity and of the Hall number. Several of the early works are not included because of missing Hall effect data [23], limited temperature range [2], non-uniform *c*-axis orientation [1], or polycrystallinity [42-44]. For measurements on untwinned crystals [8], resistivity components for transport along basal plane axes were averaged. We focus on the temperature range from 100K to 300K, where most of the data for the normal state are available. The region near $T_C$ is excluded, owing to the influence of superconducting fluctuations [15,45-47] and flux-flow pinning [48] near the superconducting transition in a magnetic field.

Basic attributes of the data are listed in table 1, including resistivity $\rho$, Hall number density $n_H$, and Hall mobility $\mu_H$ determined at T=200 K (a representative temperature at the midpoint of the temperature range under study). A histogram of the distribution of $\mu_H$ at T=200 K among the 14 specimens is shown in figure 1. Variations in mobilities among the specimens are presumed to arise from experimental errors (e.g., contact placement and sample dimensions) and sample quality (e.g., stoichiometric variations of cations and state of oxidation).

Guided by contemporaneous theoretical considerations, several authors analysed their data in terms of a quadratic temperature dependence of $\cot\theta_H$. This is equivalent to considering the inverse of the Hall mobility $\mu_H^{-1}$, which we tested for the 14 specimens in the present survey by fitting the data to a model power-law function,

$$\mu_H^{-1} = \mu_{H0}^{-1}[(T/T_0)^p + c] \,, \tag{1}$$

**Table 1.** Characteristic attributes of resistivity and Hall effect data in (a) eight crystal and (b) six film specimens of a$YBa_2Cu_3O_{7-\delta}$. Resistivity $\rho$, Hall number density $n_H = 1/e|R_H|$, and Hall mobility $\mu_H$ are given for representative temperature T=200 K. Coefficient $\mu_{H0}$, offset c, and exponent *p* are values (statistical errors) obtained by fitting Hall mobility to a power law, Eq. (1). Parameters $\sigma_{\rho/T}$ and $\sigma_{TR_H}$ are fractional standard deviations from linearity in temperature for $\rho$ and $n_H$, respectively, defined by Eq. (19).

(a) $YBa_2Cu_3O_{7-\delta}$ Crystals.

| reference | $\rho$(T=200K) ($\mu\Omega$cm) | $n_H$(T=200K) ($10^{21}$cm$^{-3}$) | $\mu_H$(T=200K) (cm$^2$/Vs) | $\mu_{H0}$ (cm$^2$/Vs) | c | p | $\sigma_{\rho/T}$ | $\sigma_{TR_H}$ |
|---|---|---|---|---|---|---|---|---|
| 3 | 159.6 | 7.74 | 5.05 | 5.44(10) | +0.07(2) | 2.15(4) | 0.0378 | 0.0247 |
| 6$^{(1)}$ | 93.6 | 9.59 | 6.95 | 4.71(19) | -0.17(3) | 1.67(5) | 0.0582 | 0.1200 |
| 6$^{(2)}$ | 96.9 | 11.61 | 5.55 | 5.00(8) | -0.09(1) | 1.67(2) | 0.0382 | 0.0919 |
| 7 | 94.6 | 11.61 | 5.67 | 5.43(7) | -0.04(1) | 1.76(2) | 0.0318 | 0.0919 |
| 8 | 104.2 | 9.26 | 6.47 | 5.76(5) | -0.11(1) | 1.67(2) | 0.0302 | 0.0422 |
| 10 | 162.0 | 7.45 | 5.17 | 5.45(56) | +0.06(9) | 2.08(20) | 0.0299 | 0.0308 |
| 12 | 139.6 | 7.85 | 5.70 | 5.24(9) | -0.08(1) | 1.72(3) | 0.0362 | 0.0714 |
| 19 | 320.7 | 5.36 | 3.63 | 3.48(9) | -0.05(2) | 1.93(7) | 0.0327 | 0.0725 |

(1) Sample 1 of Ref. [6]; (2) Sample 2 of Ref. [6].

(b) $YBa_2Cu_3O_{7-\delta}$ Films.

| reference | $\rho$(T=200K) ($\mu\Omega$cm) | $n_H$(T=200K) ($10^{21}$cm$^{-3}$) | $\mu_H$(T=200K) (cm$^2$/Vs) | $\mu_{H0}$ (cm$^2$/Vs) | c | p | $\sigma_{\rho/T}$ | $\sigma_{TR_H}$ |
|---|---|---|---|---|---|---|---|---|
| 13 | 212.4 | 5.21 | 5.64 | 5.38(1) | -0.048(2) | 1.63(3) | 0.0403 | 0.0617 |
| 14 | 153.1 | 6.88 | 5.93 | 5.57(6) | -0.051(9) | 1.64(2) | 0.0084 | 0.0780 |
| 12 | 154.4 | 6.76 | 5.97 | 5.51(7) | -0.08(1) | 1.67(2) | 0.00493 | 0.0383 |
| 15 | 82.9 | 10.60 | 7.11 | 6.75(9) | -0.06(1) | 1.46(2) | 0.0438 | 0.0848 |
| 16 | 129.2 | 7.74 | 6.24 | 6.01(32) | -0.04(4) | 1.76(9) | 0.0285 | 0.736 |
| 17 | 275.6 | 3.67 | 6.18 | 5.69(8) | -0.08(1) | 1.60(2) | 0.0163 | 0.0515 |

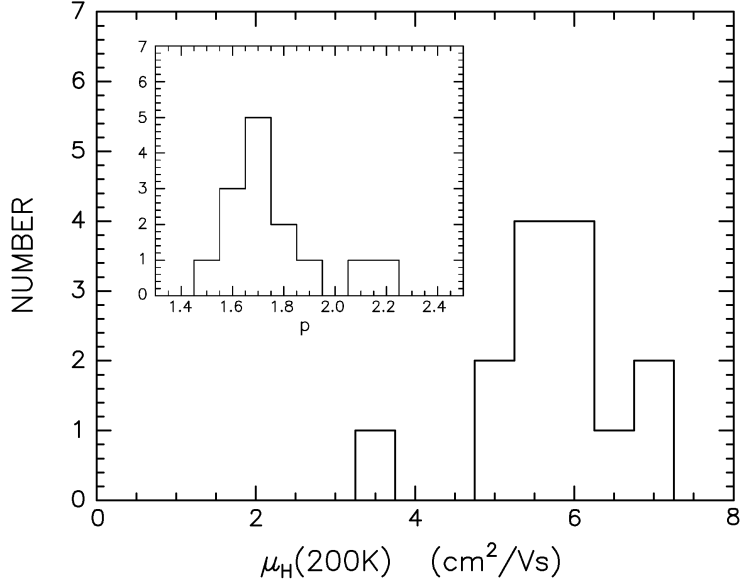

**Figure 1.** Distribution of the Hall mobility $\mu_H$ at T=200 K (bin width 0.5 cm$^2$/Vs) for the 14 surveyed specimens of YBa$_2$Cu$_3$O$_{7-\delta}$. Inset: Distribution of the exponent $p$ (bin width 0.1) obtained by fitting the temperature dependence of the inverse Hall mobility to Eq. (1).

where normalization temperature $T_0 \equiv 200$ K is introduced to yield a conventionally dimensioned coefficient $\mu_{H0}^{-1}$ and a dimensionless exponent $p$ (not restricted to integer values). The temperature independent term, denoted by the dimensionless offset parameter c, is expected to be positive for residual scattering by defects and impurities under the assumption that $\mu_H$ reflects single-band transport. The results for $\mu_{H0}$, c, and $p$ are shown in table 1. The histogram inset to figure 1 shows that the exponent $p$ varies significantly among the specimens and that the peak in the distribution is closer to 1.7 than to the integer 2, the latter being the exponent assumed by various previous authors in fitting or displaying cot $\theta_H$ vs. T$^2$ or (cot $\theta_H$)$^{1/2}$ vs. T. Since the value of $p$ is clearly sample-dependent, one cannot deduce an intrinsic fundamental value for it.

The fact that c < 0 for all but one sample (within statistical error) implies the additional presence of negatively charged carriers; there exist at least two bands where the Hall coefficient components are of opposite signs, yielding partial compensation in the result for $\mu_H$. This important finding, which was overlooked in previous work, shows that the coexistence of electrons and holes is necessary to explain the normal state resistivity and Hall effect.

## 3. Two-band model: resistivity and Hall coefficient

We show that the electrical transport properties in the a-b plane of YBa$_2$Cu$_3$O$_{7-\delta}$ in the normal state are consistent with a two-band model comprising electron and hole carriers of opposite signs. A generalized approach is adopted for the mobilities of carriers in the bands, in contrast to earlier treatments intended to explain a specific temperature dependence of cot $\theta_H$ or $\mu_H^{-1}$. Both DC and AC (far-infrared) Hall effect data are studied, yielding mutually consistent results.

The electrical resistivity $\rho$ is modeled with parallel transport by electrons and holes, where the respective resistivity components are $\rho_n$ and $\rho_p$, and given by

$$\rho = (\rho_n^{-1} + \rho_p^{-1})^{-1} . \qquad (2)$$

In the following we develop our two-band model as applied to the specific cases of the DC and AC Hall effect. The model is applied to analysis of published experimental data in section 4.

## 3.1 DC Hall effect coefficient

In our two-band model, the Hall coefficient $R_H$ can be written in terms of the component Hall coefficients of the electrons and the holes, $R_{Hn}$ and $R_{Hp}$, respectively [49]:

$$R_H = \frac{\rho_n^{-2} R_{Hn} + \rho_p^{-2} R_{Hp}}{(\rho_n^{-1} + \rho_p^{-1})^2} . \qquad (3)$$

Here, we take $R_{Hn}$ to be negative, $R_{Hp}$ to be positive, and both quantities to be temperature-independent materials constants. The applicability of Eq. (3) to describe the measured Hall coefficient clearly follows from the thesis that the Hall scattering rate is a function of the hole and electron resistivity components which determine the scattering rates for electrical transport. Transport mobilities, $\mu_T(T=200K) = 5.1 \pm 0.7$ cm$^2$/Vs, as determined from charge modulation experiments in thin films [50], are of similar magnitudes as the Hall mobilities given in table 1, although $\mu_T^{-1}(T)$ is a nearly linear function of temperature with exponent $p_T = 1.06 \pm 0.03$.

The model presented here takes into account that $\rho(T)$ and $R_H(T)$ both vary with T in the normal state and $R_H(T)$ is positive. In principle, one does not assume any particular functional form for $\rho(T)$ and $R_H(T)$, since one must recognize (see section 2) that the temperature dependences are not identical among the various specimens studied. While useful in elucidating two-band behaviour, Hall effect analysis does not convey information on carrier concentrations, given that the two Hall factors are not determined independently.

For convenience, we define the following T-dependent variables, $w = R_H/R_{Hp}$ and $u = \rho_n/\rho_p$, and the T-independent constant, $r = -R_{Hn}/R_{Hp}$. The solutions to Eqs. (2) and (3) can then be written in the forms

$$\rho_n = \rho\,(1+u) \quad \text{and} \quad \rho_p = \rho\,\frac{1+u}{u} , \qquad (4)$$

where,

$$u = \frac{w + \sqrt{w + r + rw}}{1 - w} . \qquad (5)$$

Thus a positive $R_H$ imposes the condition $u^2 > r$. Also, positive-definite solutions of Eq. (5) require that $w < 1$.

The solution of Eq. (3) for the Hall coefficient can be rendered unique by making two assumptions. First, we let the two Hall coefficient components be equal, i.e., let $r = 1$, and define a single Hall constant as $R_{H0} = -R_{Hn} = R_{Hp}$. This introduces a relatively minor loss of generality, since the quantities $(\rho_n/\rho_p)^2$ and $R_{Hn}/R_{Hp}$ in Eq. (3) are strongly correlated (approximately linearly) so that cases for $r \neq 1$ can be obtained by scaling $\rho_n/\rho_p$. It then follows from $r = 1$ that the Hall number density $n_H(T)$ is an increasing function of temperature when $\rho_n(T) > \rho_p(T) > \rho(T)$, which can be satisfied with $\rho_n(T)$ having a relatively weaker temperature dependence than $\rho_p(T)$, and possibly containing a non-zero intercept at zero temperature.

An assumption unique to this work is to employ a model function for the temperature dependence of the electronic component of the resistivity $\rho_n(T)$. The various forms for $\rho_n(T)$ examined in the data analyses include polynomials in T and several other functions modeling non-linearity in T. Functions that contain two adjustable parameters are the linear function,

$$\rho_n^I(t) = a_I + b_I t, \qquad (6)$$

and the non-linear exponential,

$$\rho_n^{II}(t) = a_{II} \exp(b_{II} t) . \qquad (7)$$

For convenience we introduce the dimensionless variable $t=T/T_0$. Non-linear forms that contain three adjustable parameters are the power law with arbitrary exponent $c_{III}$,

$$\rho_n^{III}(t) = a_{III} + b_{III} t^{c_{III}}, \qquad (8)$$

and the second order polynomial,

$$\rho_n^{IV}(t) = a_{IV} + b_{IV} t + c_{IV} t^2, \qquad (9)$$

Equation (6) is a special case of Eq. (9). Results obtained with models of Eqs. (6) – (9) are discussed in section 4. Equation (9) is of particular interest in that it includes a constant term to represent the residual resistivity, a linear-in-T component similar to the behaviour of $\rho(T)$ and $\mu_T^{-1}(T)$ measured at DC and scattering rates measured at high frequencies, plus a non-linear correction quadratic in T.

The temperature-dependent hole resistivity $\rho_p(T)$ is determined as a solution of Eq. (2):

$$\rho_p(T) = [\rho^{-1}(T) - \rho_n^{-1}(T)]^{-1} . \qquad (10)$$

It can also be evaluated from Eq. (4) as a consistency check on data analysis. This, in combination with Eqs. (2) and (3) and our model $\rho_n^j$ functions above, allow one to treat $R_{H0}$ (or Hall number density $n_{H0} \equiv 1/|e||R_{H0}|$) as a fitting constant.

## 3.2 AC Hall effect: Hall angle and scattering rates

Our two-band model can also be applied to Hall effect measurements extending into the infrared. Within this context, the Hall angle components are defined as

$$\theta_{Hp,n} = \frac{\omega_{Hp,n}}{\gamma_{p,n}}, \qquad (11)$$

where $\omega_{Hp,n} = eB/m_{p,n}$ are the Hall frequencies, $\gamma_{p,n}$ are the scattering rates, B is the magnetic field, p denotes holes, and n denotes electrons. At typical measurement fields (B < 10 T) the Hall angle in high-$T_C$ superconductors is small compared to unity, so it can be expressed as $\theta_H = R_H B/\rho$, and we have from Eqs. (2) and (3)

$$\theta_H = \frac{\rho_p^{-2} R_{Hp} + \rho_n^{-2} R_{Hn}}{\rho_p^{-1} + \rho_n^{-1}} B. \qquad (12)$$

In our model the resistivity components are expressed in terms of scattering rates as

$$\rho_{p,n} = \frac{m_{p,n}\gamma_{p,n}}{n_{p,n}e^2}, \tag{13}$$

where $m_{p,n}$ are the effective masses, and $n_{p,n}$ are the Hall carrier densities of the holes and electrons. The expression for the Hall angle can then be written in the form

$$\theta_H = \frac{\dfrac{\theta_{Hp}}{\rho_p} + \dfrac{\theta_{Hn}}{\rho_n}}{\rho_p^{-1} + \rho_n^{-1}} = \frac{\theta_{Hp}\rho_n + \theta_{Hn}\rho_p}{\rho_p + \rho_n}. \tag{14}$$

Substituting resistivities from Eq. (13) into the right hand side of Eq. (14) yields,

$$\theta_H = \frac{\dfrac{\omega_{Hp}m_n\gamma_n}{n_p\gamma_p} + \dfrac{\omega_{Hn}m_p\gamma_p}{n_n\gamma_n}}{\dfrac{m_p\gamma_p}{n_p} + \dfrac{m_n\gamma_n}{n_n}}. \tag{15}$$

Introducing two parameters, $r_M = m_n/m_p$ for the ratio of masses and $r_N = n_n/n_p$ for the ratio of carrier densities, Eq. (15) is expressed in the form

$$\theta_H = \frac{\omega_{H0}}{m_p/m_0} \frac{\dfrac{r_M\gamma_n}{r_N\gamma_p} - \dfrac{\gamma_p}{r_N\gamma_n}}{\gamma_p + \dfrac{r_M\gamma_n}{r_N}}, \tag{16}$$

where $\omega_{H0} = eB/m_0$ and $m_0$ is the electron rest mass, noting that the sign of $\omega_{Hn}$ is negative for the electron component (charge is $-e$).

The complex Hall angle at finite frequency $\omega$ is then obtained (from drift-diffusion theory) by making the substitution,

$$\gamma_{p,n} \to \gamma_{p,n} - i\omega. \tag{17}$$

Note that for a single band, the complex Hall angle reduces to the simple Lorentzian form of Drude theory,

$$\theta_H = \omega_H/(\gamma - i\omega). \tag{18}$$

## 4. Analysis and discussion

We preface our analysis of the DC Hall effect data with quantitative evaluations of the temperature dependences of the published experimental data for $YBa_2Cu_3O_{7-\delta}$ that were discussed in section 2. Fractional standard deviations from constancy of $\rho/T$, denoted as $\sigma_{\rho/T}$, and deviations from constancy of $TR_H$, denoted as $\sigma_{TR_H}$, provide quantitative measures of deviations from linearity in the temperature dependences of $\rho$ and $1/R_H$, respectively. The deviations are defined in terms of normalized variances of a function F (either as $F = \rho/T$ or $F = TR_H$) as

$$\sigma_F^2 = \frac{\frac{1}{n}\sum_{i=1}^{n}(F_i - \overline{F})^2}{\overline{F}^2}, \tag{19}$$

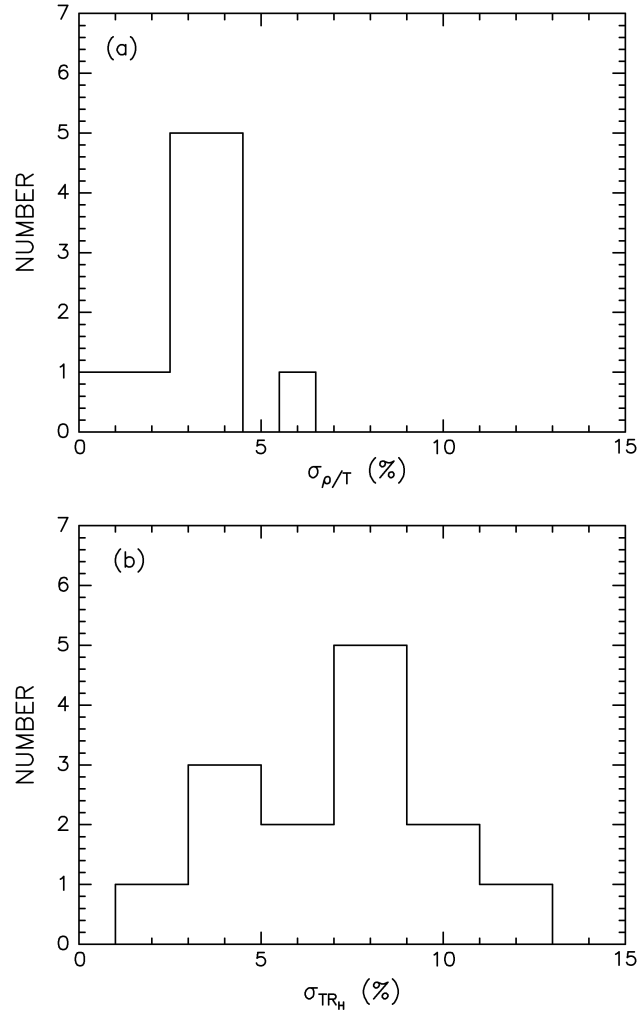

**Figure 2.** Distributions of fractional standard deviations (%) from linearity in (a) resistivity $\sigma_{\rho/T}$ (bin width 1%) and (b) Hall coefficient $\sigma_{TR_H}$ (bin width 2%) for the 14 specimens surveyed in this work (from table 1).

where i is the index of a set of n data points $\{T_i, \rho_i,$ and $R_{Hi}\}$ and

$$\overline{F} = \frac{1}{n}\sum_{i=1}^{n} F_i . \tag{20}$$

The results are shown in table 1. Variations among the specimens in the form of histogram distributions of the linearity deviations, $\sigma_{\rho/T}$ and $\sigma_{TR_H}$, expressed as percentages, are shown in figure 2. Note that for some specimens, $\rho(T)$ is nearly proportional to temperature, i.e., $\sigma_{\rho/T} \approx 0$ and $\rho(T\to 0) \approx 0$, implying negligibly small impurity scattering. The Hall number generally shows greater deviations from linear-in-T than the resistivity.

## *4.1 DC Hall effect*

In application of our two-band model for the Hall effect, we analysed data for the Hall number density $n_H(T)$ for the various specimens in our survey according to the model function of Eq. (3), evaluated as $1/|e||R_H(T)|$. The electronic component of the resistivity is determined by one of the model functions $\rho_n^j(T)$ in Eqs. (6) − (9) for j = I … IV. The hole component of the resistivity is determined according to Eq. (10). Results for each of the four model functions were determined by non-linear regression fits to the data for the Hall effect and resistivity as reported for each of the 14 specimens of $YBa_2Cu_3O_{7-\delta}$ [3, 6-8, 10, 12-17, 19]. The Hall number density parameter $n_{H0}$, as well as the parameters $a_j$, $b_j$, (and $c_j$) were treated as the three (or four) adjustable parameters.

Functions for j = II, III, and IV that contain non-linearity in the temperature dependence produce fits with statistical significances greater than the linear function for j = I. Of the three non-linear functions, the polynomial of Eq. (9) for j = IV fits the theoretical to experimental $n_H(T)$ with the best statistical significance, yielding rms normalized deviations $\sigma_{fit}$ averaging (1.3 ± 0.7) % (error bar denotes variation of $\sigma_{fit}$ among specimens). The non-linear models for $\rho_n(T)$ yield a common result: extrapolation of $\rho_n(T)$ to T=0 indicates the presence of a residual resistivity in the electronic component. Results obtained with the polynomial theoretical function j = IV, when averaged over the 14 specimens, are $\langle\rho_n(T\to 0)\rangle = 0.27 \pm 0.16$ mΩcm and $\langle n_{H0}\rangle = \langle 1/e|R_{H0}|\rangle = 2.3 \pm 1.0 \times 10^{21}$ cm$^{-3}$. As indicated by the error bars, there are substantial variations in the results from specimen to specimen, owing to variations in the magnitudes and temperature dependences of the resistivities and Hall coefficients. We discuss in section 5 that the results for the residual electronic component, $\rho_n(0)$, are consistent with the effective residual resistivity inferred from thermal conductivity data ($\rho_{eff}(0) = 0.129$ mΩcm).

Averaging the survey results implicitly assumes that the data for all specimens have equal statistical significance and merit. To quantitatively assess that this is the case we determined a figure of merit defined as $M = (\sigma_{\rho/T}\sigma_{TR_H}\sigma_{fit})^{1/3}$, which is the geometric mean of three dimensionless standard deviation attributes of the data for each specimen, where $\sigma_{\rho/T}$ expresses the linearity in the temperature dependence of $\rho(T)$, $\sigma_{TR_H}$ expresses the linearity in $1/R_H(T)$, and $\sigma_{fit}$ expresses the accuracy of the Hall effect model in fitting $n_H(T)$. In applying this criterion the average and standard deviation of the figure of merit over the 14 specimens is $\langle M\rangle = (1.85 \pm 0.69)$ %. Statistical significance was determined by examining correlations between $M$ and the various fitting parameters. Linear-correlation coefficients $R^2$ characterizing the variation of $\rho_n(T\to 0)$ with $M$ are 0.24 for the crystal specimens and <0.01 for the film specimens. For the variation of $n_{H0}$ with $M$, $R^2$ is 0.07 for the crystal specimens and 0.08 for the film specimens. These low values of $R^2$ show that the results of the Hall effect fits can be considered as randomly distributed among both the crystal and thin film specimens.

**Table 2.** Values of parameters obtained by fitting two-band Hall effect model to composite $YBa_2Cu_3O_{7-\delta}$ data for 8 crystal and 6 film specimens: $a_{IV} \equiv \rho_e(T\to 0)$, $b_{IV}$, and $c_{IV}$ determine the electron component of resisitivity, Eq. (9), and $n_{H0}$ is the Hall number density parameter.

| Specimens Type | $a_{IV}$ ($\mu\Omega$cm) | $b_{IV}$ | $c_{IV}$ | $n_{H0}$ ($10^{21}$cm$^{-3}$) |
|---|---|---|---|---|
| Crystals | 220 ± 12 | 165 ± 13 | 77 ± 5 | 2.40 ± 0.05 |
| Films | 217 ± 17 | 185 ± 17 | 59 ± 7 | 1.82 ± 0.07 |

With the above assurance that all specimens yield resistivity and Hall effect data of comparable significance, we generalize the survey analysis by averaging the data for all of the crystal specimens and separately for all of the film specimens [51]. Composite data thus obtained essentially yield more reliable experimental functions for the resistivity $\rho(T)$ and Hall number density $n_H(T)$ for electrical transport in the a-b plane of $YBa_2Cu_3O_{7-\delta}$ in the normal state.

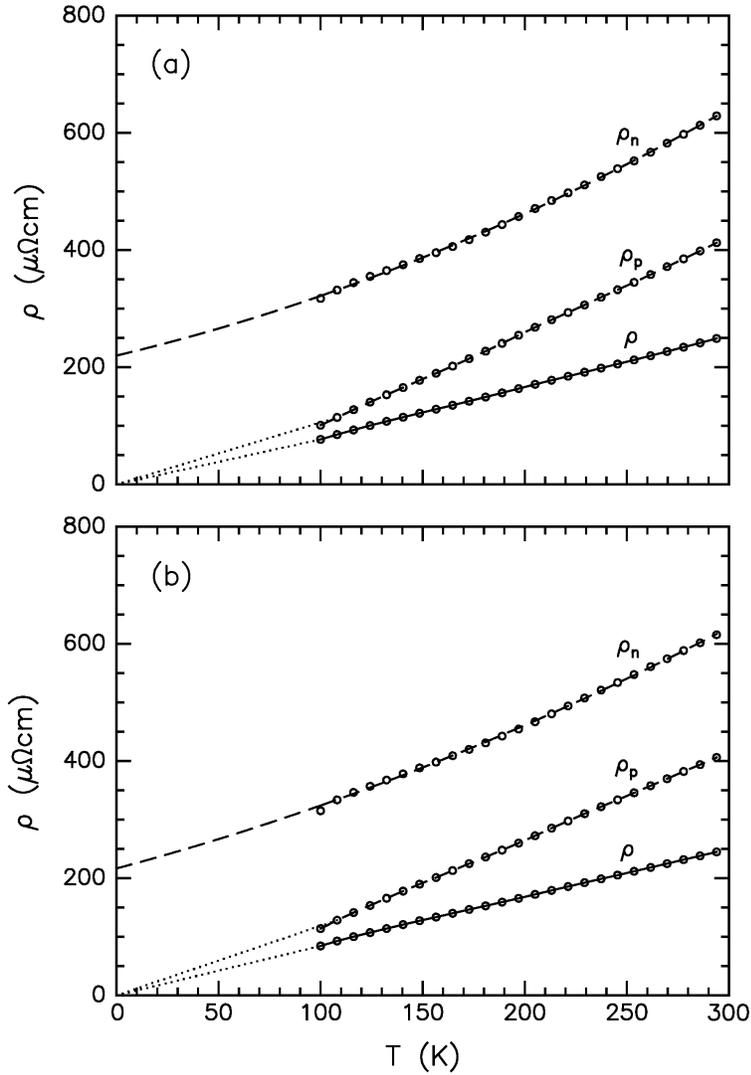

**Figure 3.** Temperature dependences of composite measured resistivity $\rho$ (points and solid curve), and fitted electron $\rho_n$ and hole $\rho_p$ components (points and dashed curves). Dotted lines are guides to the eye connecting an undetermined interval between the origin and the superconducting transition. (a) Crystal specimens; (b) film specimens.

The results for the parameters $a_{IV}$, $b_{IV}$, $c_{IV}$ and $n_{H0}$ obtained by fitting the composite data to the two-band model using the polynomial function of Eq. (9) for $\rho_n(T)$ are presented in table 2. Figure 3 shows the temperature dependences of the resistivity $\rho$ and its decomposition into components $\rho_n$ and $\rho_p$ for the (a) crystal and (b) film specimens. The dashed curves represent a fit of Eq. (9) to the data assuming the parameters of table 2. The dotted lines spanning the region between $T \sim T_C$ and $T=0$ for $\rho(T)$ and $\rho_p(T)$ (where their temperature dependences are undetermined) are linearly interpolated guides to the eye, representing only that the behaviours observed for $T>T_C$ point to zero intercepts at $T=0$. Figure 4 shows the corresponding Hall number density $n_H$. For both specimen types, the normalized standard deviation $\sigma_{fit}$ is 1% or less.

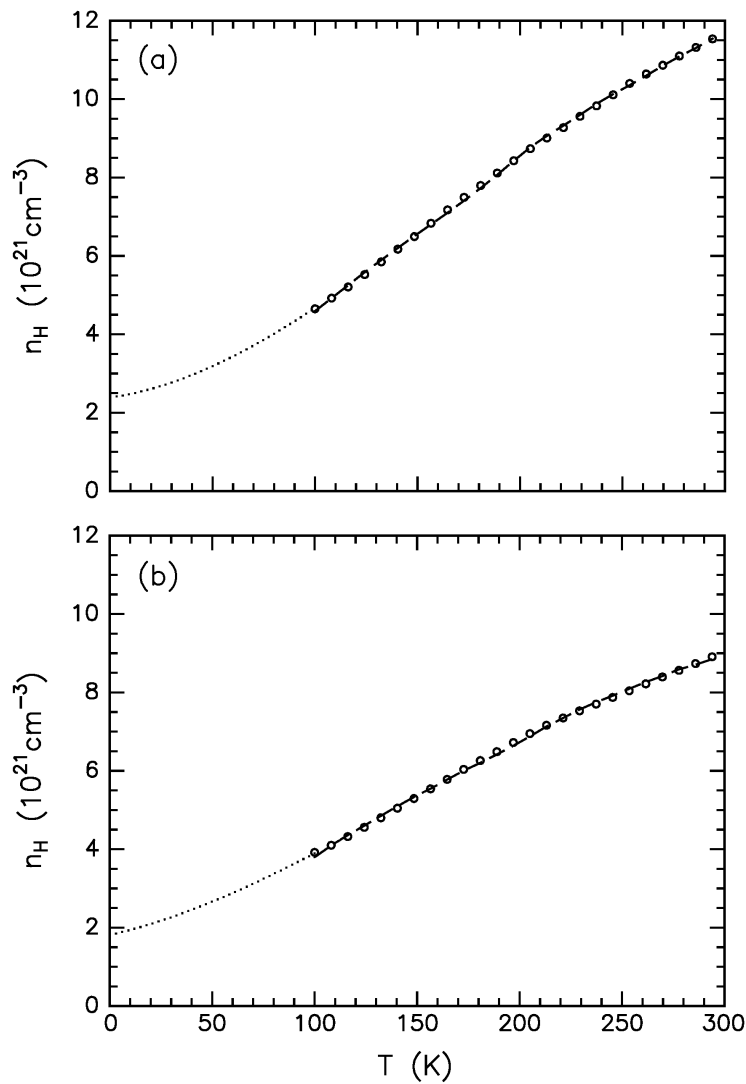

**Figure 4.** Temperature dependence of composite Hall number density (points) and fitted function (dashed curve). Dotted curved is interpolation from near $T_C$ to $T=0$. (a) Crystal specimens; (b) film specimens.

Interestingly, figure 3 shows that the composite resistivities $\rho(T)$ for the crystals and films are similar to one another. Consequently, values of $\rho_n(T\rightarrow 0)$ obtained for the crystal and film composite data differ by only 1 standard deviation, averaging $0.22 \pm 0.02$ mΩcm. However, there is a significant difference between the Hall number densities, where $n_H$ is about 30% greater for the crystals, when compared to films. This carries over into the results for the parameter $n_{H0}$, which is 32% larger for the crystals. For comparison, the fitting errors in $n_{H0}$ are 4% or less. Even though differences in the Hall number densities reveal a systematic difference between bulk-crystal and oriented-film materials, apparently reflecting their quite different growth methods, the two-band electron-hole model fits the two material forms of $YBa_2Cu_3O_{7-\delta}$ with comparable statistical significance.

In figure 3 the points for $\rho_n$ represent the fitted function of Eq. (9) and the points for $\rho_p$ are computed from Eq. (10) in terms of the data for $\rho$ and the fit for $\rho_n$. The points for $\rho$ are data, which are connected by a solid curve. Dashed curves through the points for $\rho_n$ and $\rho_p$ are the functions given in Eq. (4), where Eq. (5) is solved for u using $w = n_{H0}/n_H$ and $r = 1$, and where $n_H$ is temperature-dependent data and $n_{H0}$ is the fitting constant. The dashed curve for $\rho_n$ plots the fitting function from near $T_C$ to T=0. Thus the points are directly determined from the resistivities, i.e., data for $\rho$ and fit for $\rho_n$, while the dashed curves are functions of data, i.e., $\rho$ and $n_H$, and one fitted parameter, $n_{H0}$. Dotted lines connect points for $\rho$ and $\rho_p$ to the origin as guides to the eye, noting that normal-state resistivities were not measured for $T < T_C$, leaving $\rho_p$ in the superconducting state undetermined.

The points in figure 4 are the composite data for $n_H$ while the dashed curve through the data is the fitted function calculated from Eq. (3) using the expressions of Eqs. (9) and (10) for the resistivity components. The dotted curve extended to T=0 is a guide to the eye that connects to $n_H(T\rightarrow 0) = n_{H0}$, which is the value obtained from the fitting results that indicate $\rho_n(0) \gg \rho_p(T\rightarrow 0)$. Finite $T = 0$ intercept is consistent with the theoretical prediction $\cot\theta_H(T\rightarrow 0) \rightarrow 0$ of Ref. [39], since the plots in figure 3 suggest $\rho_p(T\rightarrow 0) \rightarrow 0$.

Standard deviation parameters for linearity in temperature dependence were also calculated for the composite data. For $\sigma_{\rho/T}$ the results are 1.8% and 1.2% for the crystals and films, respectively; for $\sigma_{TR_H}$ they are 2.0% and 1.6%, respectively. For the temperature dependence of Hall mobility fitted according to Eq. (1) the values (statistical errors) of the parameters obtained respectively for the crystals and films are: $\mu_{H0} = 4.095(5)$ and $5.07(2)$ cm$^2$/Vs; $c = -0.072(3)$ and $-0.072(3)$; and $p = 1.749(6)$ and $1.587(6)$. Note that the negative offset c is the same for the two specimen types.

## *4.2 AC Hall effect*

Previous studies of the Hall effect in high-$T_C$ superconductors in the infrared and far infrared discovered that application of Drude relaxation theory for metals, where $\theta_H(\omega) = \omega_H/(\gamma_H - i\omega)$, leads to inconsistencies, notably that $\omega_H$ and $\gamma_H$ turn out to vary with $\omega$ [52-54]. While various alternative models were proposed, the approach we take here is to conduct a re-analysis of experimental results for the complex Hall angle $\theta_H(\omega)$ in terms of the two-band model developed in section 3.2. Unlike the case for DC studies discussed in the previous section 3.1, no meaningful redundancy in measurements of the AC Hall effect is available. We therefore examine two representative cases, corresponding to the frequency and temperature dependence of the real and imaginary parts of $\theta_H(\omega)$. Since thin films or peeled crystals are required to produce adequate optical transmittance signals, there is some concern that specimens for AC measurements may not be equivalent to the more bulk-like specimens used in typical DC transport experiments (e.g., sample quality characterizations like $T_C$ and resistivity are not always given).

**Table 3.** Parameters determined from fitting electron-hole AC Hall effect model to YBa$_2$Cu$_3$O$_{7-\delta}$ frequency-dependent far-IR Hall angle data (at two temperatures) [53]. Fitted parameters for hole mass $m_p/m_0$, electron-hole mass ratio $r_M$, hole scattering rate $\gamma_p$, and electron scattering rate $\gamma_n$, are shown in bold font. Ratios of component resistivities, $\rho_p$ for holes and $\rho_n$ for electrons, and total resistivity $\rho$ are derived from fitted parameters.

| Parameter | T=95K | T=190K |
|---|---|---|
| **$m_p/m_0$** | 14.1 | |
| **$r_M$** | 1.30 | |
| **$\gamma_p$ (cm$^{-1}$)** | **112** | **200** |
| **$\gamma_n$ (cm$^{-1}$)** | **238** | **230** |
| $\rho_n/\rho_p$ | 2.77 | 1.49 |
| $\rho_p$(T=190K) / $\rho_p$(95K) | 1.79 | |
| $\rho_n$(T=190K) / $\rho_n$(95K) | 0.96 | |
| $\rho$(T=190K) / $\rho$(95K) | 1.46 | |

### *4.2.1 Frequency dependence*.

The complex Hall angle in a 50-nm YBa$_2$Cu$_3$O$_{7-\delta}$ film, which was determined from measurements of Faraday rotation in the far infrared, were plotted in figure 1 of Ref. [53] as functions of frequency ($\omega = 20 - 250$ cm$^{-1}$) at four temperatures in the normal state (T = 95, 120, 150, and 190 K). The twinned film was grown on a Si substrate over a 10-nm buffer layer (T$_C$ = 89 K was reported for a similarly prepared sample in an earlier work [55]). Data were taken for B = 8 T for which $\omega_{H0}$ = 46.934 cm$^{-1}$ in Eq. (16). Experimental traces at the lowest and highest temperatures (95 K and 190 K) were sampled at discrete points and converted into digital form for analysis. Spectra thus obtained for the real and imaginary parts of $\theta_H(\omega)$ at the two temperatures were fitted simultaneously assuming the function of Eq. (16) as modified by the complex transformation of Eq. (17). As in the DC analysis and to reduce parameter correlation effects, we fix $r_N = 1$, i.e., take carrier concentrations to be equal; moreover, allowing $r_N$ to be a variable was found to yield a value already close to unity, $r_N = 1.16$. The number of fitting parameters, which equals six, are $m_p/m_0$, $r_M$, and $\gamma_{p,n}$ (corresponding to two temperatures and two carrier types). A Simplex algorithm was used to vary values of the parameters that minimize the sum of the squares of the difference between the data points and model for the real and imaginary part of the complex function $\theta_H(\omega)$, a procedure that minimizes the variance $\sigma_{Fit}^2$.

Results for the fitting parameters are shown in table 3. Also tabulated are various ratios of resistivities derived from the fitting parameters according to Eq. (13) and the ratio

$$\frac{\rho_n}{\rho_p} = \frac{r_M \gamma_n}{r_N \gamma_p} \quad . \tag{21}$$

Figure 5 shows the digitized data points and the fitted function (extrapolated to $\omega$=0) for the two components of $\theta_H(\omega)$ at the two temperatures.

It is clear that Eq. (16) gives an excellent description of the data, with rms fitting error given by $\sigma_{Fit}$ = 0.363 radian. Significantly, this good fit is obtained with masses for the electrons and holes that remain independent of both frequency and temperature, i.e., the Hall frequencies $\omega_{Hp}$ and $\omega_{Hn}$ are material constants of the film. The electrons are found to be 30% heavier than the holes. The hole effective mass of about 14 $m_0$ appears rather large, when compared ~3 times the hole-band mass obtained from infrared studies on high-quality untwinned single crystals [56]. The value and error for $m_h$ are uncertain owing to substantial noise in the data and non-optimal sample quality, e.g., depressed T$_C$ (earlier work obtained 6.6 $m_0$ from an analysis considering only holes [55]).

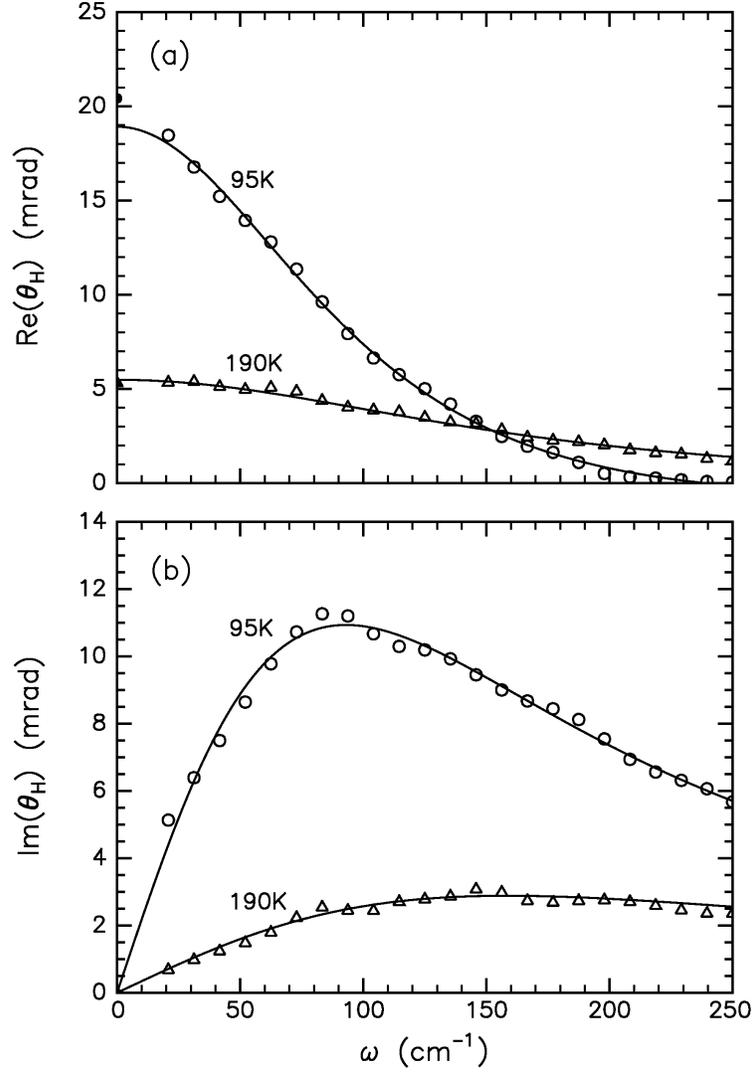

**Figure 5.** Real (a) and imaginary (b) parts of the Hall angle vs. frequency for a $YBa_2Cu_3O_{7-\delta}$ film at temperatures indicated [53]. Open symbols: far-IR data; filled symbols [on $\omega=0$ axis of panel (a)]: DC data. Curves: fitted function Eq. (16) with parameters given in table 3.

As expected, the hole scattering rate $\gamma_p$, hole resistivity $\rho_p$, and total resistivity $\rho$ increase with temperature. In contrast, however, $\gamma_n$ is found to decrease slightly with temperature, indicative of non-metallic behaviour.

The theoretical curves in figure 5 have been extrapolated to $\omega = 0$. For comparison we show as filled symbols results for $\theta_H$ from DC measurements (from figure 1 of Ref. [53]). There is a notable systematic difference between theory at $\omega = 0$ and the DC point for the T = 95 K case. Since $Re\{\theta_H(\omega)\}$ theoretically approaches $\omega = 0$ with zero slope, the discrepancy probably indicates the level of experimental uncertainty or noise in the data.

### *4.2.2 Temperature dependence.*

The real and imaginary parts of $\theta_H(\omega)$ were measured at B = 1 T and $\omega = 84$ cm$^{-1}$ ($\omega_{H0} = 5.8668$ cm$^{-1}$) over the temperature range 100 – 293K for a 100-nm $Bi_2Sr_2CaCu_2O_{8+\delta}$ peeled crystal and plotted as figure 3 in Ref. [54]. As in the treatment of previous cases, the data were digitized and $r_N = 1$ was taken in the analysis. Polynomials in temperature were used to model the temperature dependences of the scattering rates,

$$\gamma_\tau = \sum_{k=0}^{N} g_{\tau k} t^k, \tag{22}$$

where $\tau = p$ or $n$, $t = T/T_0$ and the $g_{\tau k}$ are fitting parameters, and where $T_0 = 200$ K is the normalization temperature. Second order polynomials (N=2) were found to be sufficiently robust to capture non-linearity in temperature dependence. The real and imaginary parts of $\theta_H$ were fitted simultaneously with the function of Eq. (16) as modified by the complex transformation of Eq. (17) using a procedure analogous to that of the previous section. The total number of fitting parameters in this case equals eight. Values of the fitting parameters are shown in table 4. The standard deviation in the fit is $\sigma_{Fit} = 0.0144$ mrad.

**Table 4**. Parameters determined from AC Hall effect model fit to temperature dependence of far-IR Hall angle $Bi_2Sr_2CaCu_2O_{8+\delta}$ data [54]. (a) Hole mass and electron/hole mass ratio. (b) Polynomial coefficients determining hole and electron scattering rates according to Eq. (22).

(a)

| Parameter | Value |
|---|---|
| $m_p/m_0$ | 6.97 |
| $m_n/m_p$ | 1.92 |

(b)

| Parameters | Values (cm$^{-1}$) | |
|---|---|---|
| [ $\gamma_{p,n}$ ] | $\tau = p$ | $\tau = n$ |
| $g_{\tau 0}$ | 4.4 | ~ 0 |
| $g_{\tau 1}$ | 270 | 312 |
| $g_{\tau 2}$ | 74.2 | ~ 0 |

A Hall scattering rate can be defined by the relation $\gamma_H = eB/\tilde{m}\,\theta_H$, where $\tilde{m}$ is an effective mass, which for the purpose of illustration (the scale of $\gamma_H$ being undetermined) is taken to be the sum of the hole and electron masses (this provides a prescription for dealing with the two masses). Thus we arrive at the definition

$$\gamma_H = \frac{1}{1+r_M} \cdot \frac{\gamma_p + r_M \gamma_n}{r_M \gamma_n / \gamma_p - \gamma_p / \gamma_n} \;. \tag{23}$$

The digitized data along with the fitted function for the real and imaginary parts of $\theta_H$ are plotted in figure 6. Fitted scattering rates are shown as solid curves in figure 7. For reference the rate $\gamma_H$ determined from Eq. (23) is shown as the dashed curve. Transport scattering rates [57] given in table 1 of Ref. [54] at three temperatures are shown by the symbols, which are connected by a dotted curve to guide the eye.

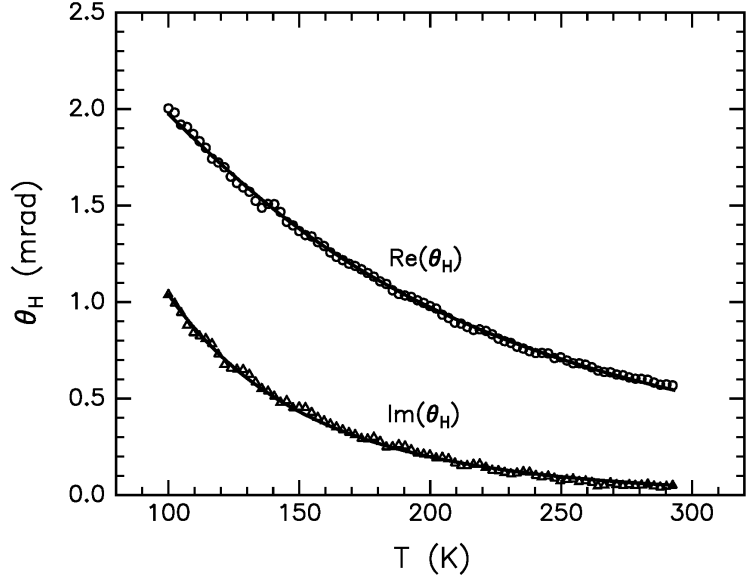

**Figure 6.** Temperature dependence of real and imaginary parts of the far-IR Hall angle for a $Bi_2Sr_2CaCu_2O_{8+\delta}$ crystal [54]. Symbols: data; curves: fitted function Eq. (16) with parameters given in table 4.

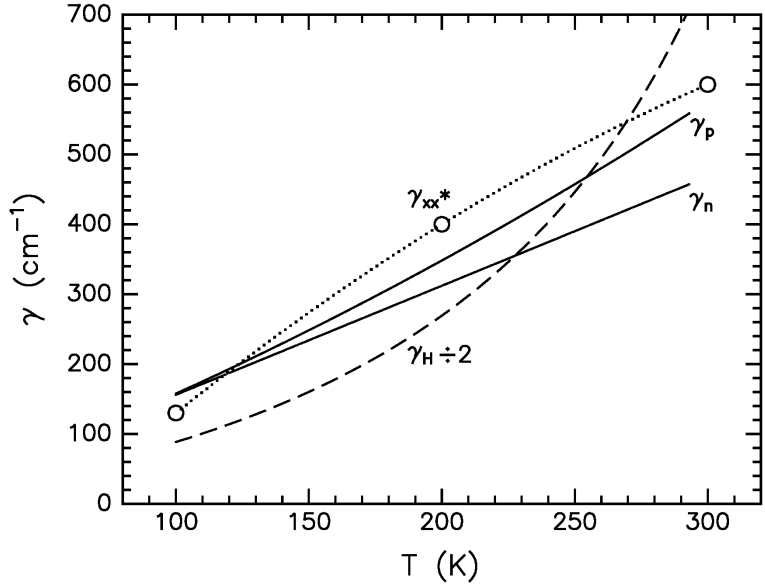

**Figure 7.** Temperature dependence of the hole ($\gamma_p$), electron ($\gamma_n$), and Hall ($\gamma_H$) scattering rates (Eq. 23) from far-IR data for a $Bi_2Sr_2CaCu_2O_{8+\delta}$ crystal [54]. Points (dotted curve as guide to eye) are transport scattering rates ($\gamma_{xx}^*$) [54,57].

We find that Eqs. (16) and (22) provide good fits to the temperature dependence of the complex Hall angle, yielding a hole mass of about 7 $m_0$ for this $Bi_2Sr_2CaCu_2O_{8+\delta}$ sample. As in the case of the $YBa_2Cu_3O_{7-\delta}$ film, the mass of the electrons exceeds that of the holes. The fitted scattering rates increase with temperature, with the electron component being nearly linear. We note that the scattering rates for the electrons and the holes are comparable to the transport scattering rates determined independently [54,57], a finding that supports our use of Eqs. (12) and (13) to model the Hall angle and resistivities in terms of transport scattering rates.

## 5. Superconducting state

One of the puzzles of high-$T_C$ superconductivity is that the residual ac conductivity at low temperature in the limit of zero frequency (e.g., extrapolated from $\sigma(\omega) \sim 2000 - 6000$ $\Omega^{-1}cm^{-1}$, $\omega \leq 200$ $cm^{-1}$) is inexplicable for moderate impurity scattering (within the context of either *d*-wave or *s*-wave pairing scenarios), and suggests the presence of a component with substantial pair breaking in the superconducting state [58]. Allowing for coexisting electron and hole bands can resolve this apparent enigma. Our analysis of DC transport in $YBa_2Cu_3O_{7-\delta}$ indicates a residual resistivity in the electronic band component, obtained from the extrapolated temperature dependence in the normal state, $\langle\rho_n(T\rightarrow 0)\rangle = 0.27 \pm 0.16$ m$\Omega$cm, as determined from data on 14 specimens, which corresponds to a residual conductivity $\sigma_n(0) \sim 2300 - 9100$ $\Omega^{-1}cm^{-1}$. The extrapolated residual DC conductivity of the electrons is thus of the same order of magnitude as the residual electrodynamic conductivity at low frequency. It is therefore elucidating to interpret this finding in terms of some known properties of the superconducting state in the limit $T\rightarrow 0$.

Harshman and Dow [59] have suggested that offsets in electronic thermal conductivity and excess electronic specific heat (terms linear in temperature T) observed in high-$T_C$ superconductors at low temperatures are evidence for a pool of electronic carriers that have non-superconducting properties (i.e., are normal-like). For numerous high-$T_C$ compounds the excess specific heat $\Delta C$ obeys the form $\gamma_0 T$ at low temperature [60]. In untwined $YBa_2Cu_3O_{7-\delta}$ single crystals $\gamma_0$ increases with $\delta$, reaching about 5 % of the normal-state coefficient $\gamma_{el}$ at optimal doping, $\delta=0.05$ (10% of $\gamma_{el}$ in some twinned crystals); and in an applied magnetic field H, the excess specific heat varies as $\Delta C = \gamma_0 T + AH^{\frac{1}{2}}T$ [61]. Having noted that the coefficient *A* does not depend on $\gamma_0$, Wen et al. have argued that the $\gamma_0 T$ term is related to an electronic inhomogeneity rather than a noded superconducting gap with small impurity scattering [62]. Thus it is consistent to associate $\gamma_0$ with a normal-like component within the superconducting phase. Considering that analyses of bond valence sum and charge conservation indicate negative charge for the $CuO_2$ layers [63][64], one may deduce from the increase in $\gamma_0$ with $\delta$ (increase in net negative charge) that the excess carriers are electrons, i.e., carry negative sign. A doping-dependent $\gamma_0$ further suggests the electronic component contains some three-dimensional character, given that the electronic density of states varies with carrier density in three-dimensional systems (but not in two-dimensional systems) [65].

An excess electronic thermal conductivity $\kappa_e$ at low temperatures, also linear in temperature, has been determined for high-$T_C$ cuprates based on Y, (La,Sr) or on (Bi,Sr) [66]. This anomaly in thermal conductivity prompted early speculation on the existence of normal carriers at low temperature, posing it as a general characteristic of high-$T_C$ superconductors [67]. This deduction remains logical, given the variability of $\kappa_e/T$ among compounds, which was found to contradict a universality prediction of *d*-wave pairing in the presence of moderate disorder [66]. While normal excitations in the superconducting state are shunted by supercurrents, they do carry entropy and thus contribute to the thermal conductivity and dissipation at finite frequency [68]. Therefore, the electronic thermal conductivity in the conventional superconducting state theoretically contains the term $f LT/\rho_0$, which is an application of the Wiedemann-Franz relation, where *f* is the normal fraction, $\rho_0$ is the residual resistivity, and $L=(\pi k_B/e)^2/3$ is the Lorentz number [68]. While *f* ideally vanishes in the limit of zero temperature, a finite *f* indicates the presence of normal-like excitations. A classic example is entropy transport in the presence of fluxons [69]. We note that the Wiedemann-Franz relation between $\kappa_e/T$ and electrical resistivity $\rho$ was found to be fairly well obeyed for $Bi_{2+x}Sr_{2-x}CuO_{6-\delta}$ (with superconductivity quenched by a strong magnetic field) [70]. In the case of optimally-doped $YBa_2Cu_3O_{6.9}$ an excess thermal conductivity given by $\kappa_e/T = 1.9\times 10^{-4}$ W/cm $K^2$ was measured in the superconducting state at low temperature [71]. Since $\rho_0$ and *f* are not determined independently, we define an effective residual resistivity as $\rho_{eff}(0) = \rho_0/f = LT/\kappa_e$, from which we calculate $\rho_{eff}(0) = 0.129$ mΩcm. Within the experimental uncertainty, the extrapolated residual resistivity of the electronic component found in section 4.1, $\langle\rho_n(T\rightarrow 0)\rangle = 0.27 \pm 0.16$ mΩcm, is consistent with this estimate of $\rho_{eff}(0)$.

Interpreting $\rho_{eff}(0)$ as the residual resistivity of normal carriers coexisting with the superconducting condensate at zero temperature, and applying a multilayer model with average layer spacing d = 0.584 nm (1/2 of the *c*-axis lattice parameter), one obtains a normal sheet resistance R = 2.2 kΩ. The scattering associated with R induces superconductive pair breaking in two-dimensions, parameterized as $\alpha = \frac{1}{2}\hbar \tau_p^{-1} = 4k_BT_{c0}\gamma R/\pi$, where $\gamma = 1.93 \times 10^{-4}$ Ω$^{-1}$ ($\approx h/e^2$) was determined experimentally for thin superconducting films [72]. This yields $\alpha = 0.54$ $k_BT_{C0}$, which is indicative of substantial pair breaking since it corresponds to reducing the transition temperature by a factor t = 0.53 [72,73]. Applying the 2-fluid model, we estimate the fraction of quasiparticle excitations as $t^4 = 0.08$, which is consistent with the observed 5 to 10% excess specific heat.

The above considerations of the low-temperature thermal properties, Hall Effect, and resistivity thus point to the presence of electrons, coexisting with the superconducting condensate of holes and maintained in metastable states via strong scattering.

## 5. Conclusion

The normal-state resistivity and Hall effect in $YBa_2Cu_3O_{7-\delta}$ and $Bi_2Sr_2CaCu_2O_{8+\delta}$ were shown to be successfully modeled by reducing band structures of these high-$T_C$ compounds to two components, comprising holes and electrons in coexistence. In addition to providing the explanation for temperature dependence in the Hall coefficient, the presence of electrons is also verified by the negative Hall mobility intercept, $\mu_H^{-1}(T\rightarrow 0) < 0$. The negative sign of the Hall coefficient for *c*-axis transport, in concert with strong scattering and evidence for three-dimensionality from doping dependence in the excess specific heat, are consistent with the presence of electrons.

Data for a substantial number of specimens of $YBa_2Cu_3O_{7-\delta}$ studied by various research groups were collected to create a large sample set for our study of the DC Hall effect. Individual Hall mobility data were analysed, starting with Eq. (1) to confirm the presence of at least two bands with oppositely signed Hall coefficients. These data were then combined to yield experimental forms with best available reliability for temperature dependences of the resistivity and the Hall coefficient for a-b basal plane transport in the normal state. The theoretical analysis does not impose particular forms for $\rho(T)$ and $1/R_H(T)$, which experimentally are nearly, but not precisely, linear functions of temperature. Analyses of bulk-crystal and thin-film materials yielded equivalent conclusions: the hole component $\rho_p(T\rightarrow 0)$ extrapolates to zero while the electron component $\rho_e(T\rightarrow 0)$ is large, consistent with a scattering rate capable of suppressing sustained superconductivity and agrees remarkably with the effective residual resistivity extracted from the thermal conductivity data of Ref. [71] using the Lorentz number ($\rho_{eff}(0) = 0.129$ mΩcm). Taking the Hall number densities of electrons and holes to be equal is sufficient for excellent agreement with experiment. Since Hall factors for the electrons and holes are not determined independently, Hall effect analysis is not expected to convey information on carrier concentrations that could be obtained by other means [37].

The electron-hole model was shown to account for the frequency and temperature dependence of the Hall angle in the far-infrared with constant values for the effective masses of the electrons and the holes, as required for a correct theoretical interpretation. We note perhaps fortuitous similarities (given the different material qualities of film and bulk samples) when hole effective masses $m_h$ derived from the AC Hall angle analysis of thin films, 14 $m_0$ for $YBa_2Cu_3O_{7-\delta}$ and 7 $m_0$ for the $Bi_2Sr_2CaCu_2O_{8+\delta}$, are compared to those determined in earlier analyses of the specific heat jump at $T_C$ for bulk specimens: $(12.0 \pm 2.4)$ $m_0$ and (7.8 ± 2.4) $m_0$, respectively [74]. The experimental errors involved however preclude drawing any quantitative comparisons between the Hall and thermal effective masses.

Through application of our two-band electron-hole model, we now have a clearer picture of the normal state transport. Since the anomalous temperature dependence of the Hall effect ($R_H \sim 1/T$) is generally observed, even in specimens with negligibly small impurity scattering ($\rho(T) \propto T$), we conclude that any approach predicated on impurity scattering of the hole carriers [38,39] would not achieve the general applicability of our electron-hole model. In addition to the two bands possessing opposite carrier sign, we find that the electrons are heavier than the holes and electrons experience significant scattering extrapolating into the superconducting state.

Extending our knowledge of the normal state below $T_C$, one may infer the following: Accepting that the holes are the dominant carriers of superconductivity, and that the high scattering rate of the electrons in the normal state continues below $T_C$, the electrons then become prime candidates for the origin of the non-vanishing ac conductivity [58], excess specific heat [60] and $\kappa/T$ offset in thermal conductivity [66]. While the holes are essentially confined to the basal-plane, the negative *c*-axis Hall coefficient [3], the doping dependence of the excess specific heat [61], and observed incoherency in *c*-axis transport [58] emerge as signatures of three-dimensional mobility for marginally metallic electrons. Such differentiations between the holes and the electrons, and the presence of the normal component at low temperatures, diminish the possibility of their forming bosonic pairs, as was proposed in Ref. [75].

Previously, evidence for the presence of two bands of charge carriers was also deduced from the influence of oxygen-isotope and cation substitutions on the transition temperature of $YBa_2Cu_3O_{7-\delta}$ [76]. For example, a small percentage of $Pr^{+3}$ substituted for $Ba^{+2}$ is destructive to superconductivity (e.g., substantially diminished Meissner effect), which is understandable in terms of coexisting hole and electron bands associated with BaO and $CuO_2$ structures, respectively. While this assignment differs from the popular single-carrier-type notion placing the holes in the cuprate planes, locating the holes and electrons according to Ref. [76] is rendered plausible as a result of our validation of their coexistence in the normal-state transport.

## Acknowledgements

We are grateful for the support of the U.S. Army Research Office (W911NF-05-1-0346 ARO), Physikon Research Corporation (Project No. PL-206), and the New Jersey Institute of Technology. The authors would also like to acknowledge that 2010 marks the 130[th] anniversary of the publication of Prof. Edwin H. Hall's seminal paper "On the New Action of Magnetism on a Permanent Electric Current" (i.e., the Hall effect) [77]. Publication on this paper has appeared [78].## References

[1] P. Chaudhari, R. T. Collins, P. Freitas, R. J. Gambino, J. R. Kirtley, R. H. Koch, R. B. Laibowitz, F. K. LeGoues, T. R. McGuire, T. Penney, Z. Schlesinger, A. P. Segmüller, S. Foner, and E. J. McNiff, Phys. Rev. B 36 (1987) p. 8903.

[2] A. Davidson, P. Santhanam, A. Palevski, and M. J. Brady, Phys. Rev. B 38 (1988) p. 2828.

[3] T. Penney, S. von Molnár, D. Kaiser, F. Holtzberg, and A. W. Kleinsasser, Phys. Rev. B 38 (1988) p. 2918.

[4] D. M. Eagles, Solid State Commun. 69 (1989) p. 229.

[5] A. T. Fiory and G. S. Grader, Phys. Rev. B 38 (1988) p. 9198.

[6] T. R. Chien, D. A. Brawner, Z. Z. Wang, and N. P. Ong, Phys. Rev. B 43 (1991) p. 6242.

[7] T. R. Chien, Z. Z. Wang, and N. P. Ong, Phys. Rev. Lett. 67 (1991) p. 2088.- 19 -